\documentclass[aps]{revtex4}
\usepackage{amssymb}
\usepackage{graphicx}
\usepackage{epstopdf}

\begin{document}

\title{A spectral collocation approximation for the radial-infall of a compact object into a Schwarzschild black hole}
\author{Jae-Hun Jung}
\email{jaehun@buffalo.edu}
\address{Department of Mathematics, SUNY at Buffalo\\
Buffalo, NY 14260, USA \\}

\author{Gaurav Khanna and Ian Nagle}\email{gkhanna@umassd.edu} 

\address{Department of Physics, University of Massachusetts at Dartmouth\\
North Dartmouth, MA 02747, USA \\}

\begin{abstract}
The inhomogeneous Zerilli equation is solved in time-domain numerically with 
the Chebyshev spectral collocation method to investigate a radial-infall of 
the point particle towards a Schwarzschild black hole. Singular source terms 
due to the point particle appear in the equation in the form of the Dirac 
$\delta$-function and its derivative. For the approximation of singular source 
terms, we use the direct derivative projection method proposed in \cite{Jung} 
without any regularization. The gravitational waveforms are evaluated as a 
function of time. We compare the results of the spectral collocation method 
with those of the explicit second-order central-difference method. The numerical 
results show that the spectral collocation approximation with the direct projection 
method is accurate and converges rapidly when compared with the finite-difference 
method.  
\end{abstract}

\maketitle

\section{Introduction}
The capture of a stellar mass compact object (a black hole or star) by a
supermassive black hole occurs commonly in the nuclei of most galaxies. As
the compact object spirals into the black hole, it emits gravitational
radiation which would be detectable by space borne detectors such as
NASA/ESA's LISA~\cite{lisa}. 

Because of the extreme mass-ratio, the small companion of a supermassive
black hole can be modeled as a point particle, and the problem can be
addressed using black hole perturbation theory. Moreover, as a first
approximation, we will let the point particle follow geodesics in the
space-time of the central black hole. This type of approach has been taken
by several researchers in the past using a frequency-domain 
decomposition of the master equation (for a recent review
see~\cite{glampedakis}). That approach has had many remarkable
achievements, specifically the accurate (to $10^{-4}$) determination of
the energy flux of gravitational waves for many classes of geodesic
orbits. However, for orbits of high-eccentricity and for {\em non-periodic} orbits 
(such as parabolic orbits, infalling or inspiraling orbits) frequency-domain 
computations typically result in long computation times and large errors.
It is here that a time-domain based approach would be invaluable.
Finite-differencing based, time-domain methods for such evolutions were only
recently developed and they suffer with a relatively poor accuracy of
about 1\%~\cite{euro,pranesh}. The main source of error in these methods
is the approximate modeling of a point particle, i.e. a Dirac $\delta$-function, 
on a finite-difference numerical grid. Various approaches to
this issue have been attempted, including ``regularizing'' the Dirac $\delta$-function 
using a narrow Gaussian distribution~\cite{euro} and also using
more advanced discrete-$\delta$ models~\cite{Tornberg3,pranesh} with only
limited success. 

An alternate approach to the time-domain based solution of this extreme mass-ratio 
inspiral (EMRI) problem is to use the spectral method instead of the finite-differencing
approach discussed above. It is well known that the spectral method yields 
so-called ``spectral convergence'' for the approximation of smooth problems. That 
is, the accuracy of the spectral method improves by an exponential order rather than 
an algebraic order which is the typical order for the finite difference method. 
The spectral method, however, loses its spectral accuracy when discontinuous 
or singular problems are considered. Convergence of the spectral approximation 
for discontinuous problems is only $\mathcal{O}(1)$ in its $L_\infty$ norm i.e. error is 
 $\mathcal{O}(1)$ for large N if the discontinuity is included and  $\mathcal{O}(1/N)$ if it is not. For 
the approximation of singular source terms, the spectral Galerkin approach 
yields an efficient approach because the Galerkin projection of the Dirac 
$\delta$-function to any polynomial spaces can be easily and exactly evaluated. 
However, the spectral Galerkin approximation is only limited to $\mathcal{O}(1)$ 
convergence near the singularity or the local jump discontinuity. The spectral 
collocation method also suffers from this $\mathcal{O}(1)$ convergence and this 
is well known as the Gibbs phenomenon. The Gibbs oscillations found in spectral 
approximations can also easily induce numerical instability especially for 
hyperbolic equations. A regularization of the solution or the singular source terms is 
therefore necessary to enhance accuracy away from 
the singular regions and also to maintain numerical stability. In \cite{Jung,Jung2} 
it has been shown that the direct collocation projection of the $\delta$-function 
can yield spectral accuracy for some linear PDEs even when no regularization is applied. 
The spectral accuracy obtained by this method is due to the consistent formulation 
of the given PDEs in the collocation sense. The same direct projection method for the 
spectral Galerkin method, however, does not show such good results. Although the direct 
collocation method is limited only to certain classes of PDEs, it can be applied to more 
general problems if the solutions exhibit some ``nice'' properties such as symmetry, steady-state 
behavior, linearity, etc. \cite{Jung}. The problem that we consider in this paper 
is described by the second-order harmonic equations with singular source terms 
and we will show that the direct projection method yields accurate results for this case. 

In this work, we perform a {\em proof-of-concept} computation to
demonstrate that the spectral collocation method is more accurate
and exhibits higher convergence rate when compared to a time-domain solution 
method for the EMRI problem based on the second-order finite-differencing. 
We demonstrate this using a {\em non-periodic} radial infall problem for a point particle into a 
Schwarzschild (non-rotating) black hole. The master equation to be solved is 
the inhomogeneous Zerilli equation~\cite{Zerilli} which is essentially a 
wave-equation with a potential and a complicated source term. We solve this 
equation two different ways -- using standard explicit second order finite 
difference method and using a Chebyshev spectral collocation method -- and 
compare the results.

This paper is organized as follows. In Section 2, we briefly outline the mathematical formulation (Zerilli equation) behind the astrophysical process under consideration. The mass-ratio of the system is assumed to be infinitesimally small so that the small object can be considered as a point source. In Section 3, we explain the numerical methodology used for the approximation of the Zerilli equation. The Chebyshev spectral collocation method is explained in detail there. For the approximation of singular source terms that appear in the Zerilli equation, we introduce the direct collocation projection method of the $\delta$-function and its derivative(s). Numerical comparison between the Chebyshev collocation method with the direct projection method and the Lax-Wendroff method for the approximation of a simple first-order wave equation is provided. In Section 4, numerical results of the Chebyshev spectral method are provided for the Zerilli equation. We also present the second-order finite difference approximations for comparison with the results of the spectral method. In Section 5, we summarize and suggest possible future work.  

\section{EMRI using black hole perturbation theory}

\begin{figure}[ht]
\begin{center}
\includegraphics[width=3in]{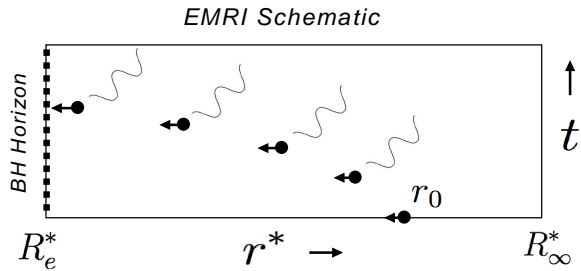}
\end{center}
\caption{The schematic illustration of the EMRI for the 1D inhomogeneous Zerilli equation Eq. (\ref{zerilli}). $R_e^*$ and $r_0$ denote the location of the event horizon and the initial location of the infalling point source. $t$ denotes the physical time. As $t$ increases, the infalling particle illustrated by the black dot with the arrow approaches the event horizon and emits the gravitational wave propagating toward $R^*_\infty$ illustrated by the solid ripples in the figure.}  
\label{schematic}
\end{figure}

In this section, we summarize the equations corresponding to an 
extreme-mass-ratio binary black hole system. We will keep 
the discussion here very brief and simply refer the reader to the more 
detailed treatment presented in reference \cite{Lousto}. 
Depicted in Figure \ref{schematic} is schematic that lays
out the basic set-up of the problem that we are addressing in this paper. 
The point particle initially starts out at a location $r_0$
and begins to ``fall'' toward the black hole event horizon as shown on the
left side of the figure. As the particle falls, it gains more speed and
radiates gravitational waves. Our numerical approach towards this problem
uses the $r^*$ coordinate and uses appropriately chosen, finite values for
the location of the boundaries. 

The central Schwarzschild black hole has a mass of $M$ and the infalling object has a mass of $m_0$. For simplicity, the mass-ratio between the two is taken to be infinitesimally small so that the small object is considered as a point source moving on a geodesic of the curved space-time due to the other, i.e. $m_0/M \rightarrow 0$. Because we are treating this problem in the context of first-order perturbation theory, the governing equation of this system is linear and the solution is the same for any mass-ratio. One solution of a particular mass-ratio can be derived from another of different mass-ratio, simply by rescaling. We assume that the point source is falling along the $z$-axis toward the hole without any angular momentum. As the point source is falling into the massive hole, gravitational waves are emitted by the system. The lowest order perturbation theory of the initial Schwarzschild black hole spacetime leads to the inhomogeneous Zerilli equation with even-parity. Such an equation describes the gravitational wave $\psi$ in $1+1$ dimension given by the following second-order wave equation with a potential and source term, 

\begin{equation}
                         \psi_{tt} = \psi_{r^*r^*} - V_l(r)\psi - S_l(r,t),
\label{zerilli}
\end{equation}
where $r^*$ is the tortoise coordinate, $V_l(r)$ the potential term and $S_l(r,t)$ the source term \cite{Lousto,Zerilli}. Note that we have set the units $c=G=1$. The tortoise coordinate mentioned above, is given by
\begin{equation}
                                r^* = r + M\ln (r/2M-1),
\end{equation}
where $r$ is the standard Schwarzschild coordinate and $r > 2M$. 
Note that tortoise coordinate $r^*$ approaches $-\infty$ as the value of $r$ approaches the Schwarzschild radius $r = 2M$. The Zerilli potential $V_l(r)$ is given by  
\begin{equation}
                               V_l(r) = f(r){{2\lambda^2(\lambda+1)r^3 + 6\lambda^2Mr^2 + 18\lambda M^2r + 18M^3}\over{r^3(\lambda r + 3M)^2}},
\end{equation}
where the function $f(r)$ is 
\begin{equation}
                               f(r) = 1 - {{2M}\over r}
\end{equation}
and the parameter $\lambda$ is 
\begin{equation}
                               \lambda = (l+2)(l-1)/2.
\end{equation}
In our case of interest, $l$-pole is only even with azimuthal angular term $m = 0$. It is noteworthy that as $r\rightarrow 2M$, $f(r)$ becomes $f(r) \rightarrow 0$ and the potential vanishes. As $r \rightarrow \infty$, the potential also vanishes. The local maximum of $V_l(r)$ exists near the Schwarzschild radius.  

Now, the source term is given by 
\begin{eqnarray}
            S_l(r,t) &=& -{{2(1-2M/r)\kappa}\over{r(\lambda+1)(\lambda r + 3M)}}
               \left[
                 -r^2(1-2M/r){1\over{2U^0}}\delta'(r - R)\right. \nonumber \\ && +
                \left. \left\{
                      {{r(\lambda+1)-M}\over{2U^0}} - {{3MU^0r(1-2M/r)^2}\over{r\lambda+3M}}
                \right\}\delta(r - R)
               \right],
\end{eqnarray}
where $\kappa$ is
$$
                \kappa = 8\pi m_0\sqrt{(2l+1)/4\pi},
$$
and the zero component of the 4-velocity of the particle $U^0$ is given by 
$$
               U^0 = \sqrt{{1 - 2M/r_0}\over{1-2M/r}},
$$
where $r_0$ is the initial position of the particle.

Singular source terms $\delta(r-R)$ and $\delta'(r-R)$ are due to the point source and its movement along $r$. Here $R$ is the coordinate of the point source at time $t$. Although the source term expression takes a complicated form, one can easily obtain some intuitive aspects of the physical system from it. For example, note that the term simply scales with $m_0$ (through the quantity defined above as $\kappa$) which is simply a reflection of our linear approximation as mentioned earlier. In addition, as expected, the expression is non-zero only at the location of the point particle ($r=R$). It is also interesting to note that, if the particle is very far from the black hole, it is the $\delta'(r-R)$ term that strongly dominates the source expression, while in the other extreme case the entire term becomes negligible. The superscript $'$ denotes the derivative with respect to $r$ not $r^*$. Since the derivative with respect to space in Eq. (\ref{zerilli}) is given in $r^*$ rather than $r$, we need to convert $r$ to $r^*$ to locate the point source at $R$. For our numerical approximation in the next section with the spectral method, $r^*$ is transformed to the unit interval $[-1,1]$ using the linear map as explained in the next section. The free fall trajectory $R(t)$ of the test particle is then given by 
\begin{eqnarray}
                          \dot{R} = -f(R) \sqrt{{2M/R-2M/r_0}\over{1 - 2M/r_0}}. 
\label{Rdot}
\end{eqnarray}
If $r_0 \rightarrow \infty$, the initial system can be regarded as the unperturbed 
Schwarzschild spacetime and the wave function $\psi$ vanishes for all $r$ at $t = 0$. 
Due to the computational limit, however, $r_0$ is only finite for the numerical approximation. 
For the current work, we assume that $m_0$ is infinitesimally small compared to the mass $M$ 
of the Schwarzschild black hole so that the initial $\psi$ is chosen to vanish. This causes 
a small discrepancy in the full solution. The initial fluctuations due to such discrepancy, 
however, leave the computational domain according to Eq. (\ref{zerilli}) as $t\rightarrow \infty$. 

\section{Numerical methodology}
In this section, we present the numerical methodology used in this paper - the spectral collocation method and the direct projection method of the $\delta$-function to solve the second-order wave equation given in Eq. (\ref{zerilli}). A numerical example for the simple advection equation is also provided. For the time integration, the TVD third-order Runge-Kutta method is used for the spectral method.

\subsection{Chebyshev spectral collocation method}
For the spatial differentiation, the Chebyshev spectral collocation method is used. Suppose that the differential operator in one dimension is given on the solution $u(x,t)$ for $x\in \Omega$ and $t>0$ as 
\begin{eqnarray}
                \mathcal{L} u = 0,
\end{eqnarray}
with the boundary conditions
\begin{eqnarray}
               \mathcal{B}^\pm u = h^\pm(t),\quad x \in \partial \Omega 
\end{eqnarray}
where $\mathcal{B}^\pm$ are the boundary operators at the domain boundary $\partial\Omega$. Let $\mathcal{P}_N$ be the projection operator applied to $u(x,t)$ onto the polynomial space $\mathcal{B}_N$ expanded by the Chebyshev polynomials of up to degree $N$. The Chebyshev spectral collocation method seeks a solution in the Chebyshev polynomial space expanded by the Chebyshev polynomials $T_l(x)$ as
\begin{eqnarray}
            u_N = \mathcal{P}_N u(x,t) = \sum_{l=0}^m {\hat u}_l (t) T_l(x),
\end{eqnarray}
where $T_l(x)$ is the Chebyshev polynomial of degree $l$ and ${\hat u}_l$ the corresponding expansion coefficient. Chebyshev polynomials are defined by the following orthogonality property
\begin{eqnarray}
                 {1\over {h_l}} \int^1_{-1} w(x) {T_l(x) T_{l'}(x)} dx = \delta_{ll'},
\end{eqnarray}
where $h_l$ is the normalization factor given by $h_l = \pi$ for $l = 0$ and $h_l = {\pi \over 2}$ for $l \in \mathbb{Z}^+$ and $w(x)$ is the Chebyshev weight function given by $w(x) = 1/\sqrt{1-x^2}$.  
The expansion coefficients $\{{\hat u}_l \}_{l=0}^m$ are found based on approximations at the collocation points defined in $\Omega$. Here $\delta_{ll'}$ is the Kronecker delta in a usual sense. Then the exact expansion coefficients ${\hat u}_l$ is obtained by 
\begin{eqnarray}
                    {\hat u}_l(t) = {1\over {h_l}} \int^1_{-1} w(x) u(x,t) T_l(x)dx. 
\end{eqnarray}
Since the collocation method seeks the solution at the particular collocation points, ${\hat u}_l(t)$ is obtained by evaluating the integral using the quadrature rules on such collocation points. As long as the notation is not confusing, we also use ${\hat u}_l$ as the approximation of the exact expansion coefficients. Since the boundary conditions are given, the commonly used collocation points are the Gauss-Lobatto quadrature points. The Chebyshev Gauss-Lobatto collocation points $x_j$ are given by 
\begin{eqnarray}
             x_j = -\cos({\pi\over N}j), \quad \forall j = 0, \cdots, N,    
\end{eqnarray}
here $N = m$ for the completeness. With this condition, we construct the approximation based on the solution at the collocation points, $\{ u(x_j,t)\}_{j=0}^N$ such as
\begin{eqnarray}
          u(x,t) = \sum_{l=0}^m {\hat u}_l(t) T_l(x) = \sum_{l=0}^m \psi_l(x) u(x_l,t) = \sum_{l=0}^m \psi_l(x) u_l,
\end{eqnarray}
where let $u_l$ denote the grid function of $u(x,t)$ at $x=x_j$ and $\psi_l(x)$ are indeed interpolating polynomials. The expansion coefficients are given by 
\begin{eqnarray}
               {\hat u}_l(t) = {2\over{c_lN}}\sum_{j = 0}^N{1\over {c_j}} u_j(t) T_l(x_j). 
\label{coefficient}
\end{eqnarray}
The interpolating polynomial $\psi_l(x)$ is a $\delta$-function in the sense that $\psi_l(x_{l'}) = \delta_{ll'}$. One may adopt $\psi_l(x)$ as an approximation of the $\delta$-function, but using $\psi_l(x)$ as a $\delta$-function for the given PDE yields only highly oscillatory solution near the singularity \cite{Jung}. The interpolating polynomials $\psi_l(x)$ on the Gauss-Lobatto-Chebyshev collocation points are given by
\begin{eqnarray}
\psi_l(x) = {{(-1)^{l+1}(1-x^2)T'_N(x)}\over{c_lN^2(x-x_l)}}, 
\end{eqnarray}
where $c_l = 1$ for $l = 1,\cdots, N-1$ and $c_l = 2$ for $l = 0, N$. By taking a derivative of $\psi_l(x)$ we obtain the derivative of $u(x,t)$ with respect to $x$ such as
\begin{eqnarray}
                       u'(x,t) = \sum_{j=0}^m \psi'_j(x) u_j,
\end{eqnarray}
where the superscript $'$ denotes the derivative with respect to $x$. Then at the collocation points we obtain
\begin{eqnarray}
                      {\bf u}' = {\bf D} \cdot {\bf u},
\end{eqnarray}
where ${\bf u} = (u_0, \cdots, u_N)^T$ and ${\bf D}$ the differentiation matrix given by the derivative of the interpolating polynomial at the collocation points as 
\begin{eqnarray}
D_{lj} = \psi'_j(x_l).
\label{Dmatrix} 
\end{eqnarray}
Thus we know that every column of $\bf D$ is the derivative of the $\delta$-function at the collocation points. The elements of $\bf D$ with the Chebyshev Gauss-Lobatto collocation points are given by \cite{Gottlieb,Hesthaven}, 
\begin{eqnarray}
                             D_{ij} = \left\{ \begin{array}{ccc}
                                       {{c_i}\over{c_j}} {{(-1)^{i+j}}\over{x_i - x_j}} &  i\ne j, \\
                                                  -{1\over 2} {{x_i}\over{1-x_i^2}}  &  i = j, \quad i \ne 0, N \\
                                                  {{2N^2+1}\over 6}           &  i = 0 \\
                                                  {{2N^2+1}\over 6}           & i = N \end{array} \right.  .
\label{equation19a}
\end{eqnarray}

The Chebyshev collocation method seeks a solution $u(x,t)$ by having the following discrete residue $R_N(x)$ vanish at the collocation points,
\begin{eqnarray}
             R_N(x) = \mathcal{L}_N u_N - 0 , \quad x = x_j, \forall j = 0, \cdots, N,
\end{eqnarray} 
where $\mathcal{L}_N = \mathcal{P}_N\mathcal{L}\mathcal{P}_N$.

\subsection{Spectral projection of singular source terms}
For singular source terms, we use the direct projection method \cite{Jung}. The direct projection method is to approximate the $\delta$-function in a consistent way with the given PDE. Suppose that we are given a PDE in one dimension 
$$
                          \mathcal{L}u = \delta(x), \quad x\in \Omega = [-1,1],
$$
where $\mathcal{L}$ = $d/dx$ and the boundary condition $u(-1) = 0$. Here the $\delta$-function in the bounded domain $\Omega$ is defined as 
\begin{eqnarray}
    \int^1_{-1} \delta(x) dx = 1, \quad \mbox{and }\quad \int^1_{-1}\delta(x) u(x) dx = u(0).  
\label{definition}
\end{eqnarray}
Then the exact solution is given by the Heaviside function $H_c(x)$ 
$$
                         H_c(x) = \left\{ \begin{array}{cc} 0, & x<c \\ 1, & x>c \end{array}\right. ,
$$
where $c = 0$ for this case. Then the direct projection method for the $\delta$-function is obtained by simply mimicking the relation
$$
                         {{d}\over {dx}} H_c(x) = \delta(x),
$$
at the collocation points such as 
\begin{eqnarray}
           \delta_N = {\bf D} \cdot {\bf H},
\label{SPdelta} 
\end{eqnarray}
where $\delta_N$ is the collocation approximation at the collocation points, $\delta_N = (\delta_N(x_0), \cdots, \delta_N(x_N))^T$, and $\bf H$ is ${\bf H}=(0, \cdots, 0, 1, \cdots, 1)^T$. For $\bf H$ the nonzero value begins at $x = x_j$ such that $x_j >c$. The consistent formula of $\delta_N$ does not necessarily satisfy the definition of the $\delta$-function in Eq. (\ref{definition}). Suppose that $u(x) = |sgn(x)|$. Then by definition at each collocation point $x_j$,
$$
             {d\over{dx}} \int^x_{-1} \delta(s)u(s) ds = \delta(x)u(x) = 0,  \quad \mbox{for } x_j, \forall j = 0, \cdots, N. 
$$ 
At $x_j$ the consistent formula of $\delta_N$ yields
$$
                       \delta(x)u(x)|_{x_j} \rightarrow ({\bf D}\cdot{\bf H}_N)_j u(x_j) = ({\bf D}\cdot{\bf H})_j \ne 0, \mbox{for } x_j,  \forall j = 0, \cdots, N.  
$$
For the time dependent problem such as the Zerilli equation considered in the next section, the source term is also a function of time. In this case, the parameter $c$ for $H_c(x)$ is a function of time, i.e. $c = c(t)$. If $c(t)$ coincides with one of the collocation points, $x_j$ at a certain time $t_p$, the definition of the Heaviside function $H_c(x)$ at $x_j$ and $t_p$ can be either $H_{c(t_p)}(x_j) = 1$ or $H_{c(t_p)}(x_j) = {1\over 2}$. Both definitions yield the same results.

\subsection{Numerical examples}
To demonstrate the accuracy obtained by the direct projection method for the 1D wave equation with the singular source term, we consider the following advection equation 
\begin{eqnarray}
                         u_t + u_x = \delta(x), \quad x\in [-1,1], \quad t>0,
\label{example1}
\end{eqnarray}
with the initial condition $u(x,0) = -\sin(\pi x)$ and $u(-1,t) = -\sin(\pi(-1-t))$ for $t>0$. The exact solution of Eq. (\ref{example1}) is then given by 
$$
                 u(x,t) = -\sin(\pi(x-t)) + \int^x_{-1} \delta(x) dx. 
$$
As the exact solution implies, the solution $u(x,t)$ is discontinuous at $x = 0$. The jump magnitude $[u]$ at $x = 0$ is $1$ for any time $t$,
$$
     [u]|_{x=0} := u(0^+) - u(0^-) =  \int^{0^+}_{-1} \delta(x)dx - \int^{0^-}_{-1} \delta(x) dx = 1. 
$$
For the numerical experiment, we consider $N = odd$. Then the exact solution $u(x,t)$ at the Gauss-Lobatto collocation points $x=x_i$ and at $t$ are given by 
$$
                u(x_i,t) = \left\{ \begin{array}{cc} -\sin(\pi(x_i-t)) & i \in [0,\cdots, {{N+1}\over 2}] \\
                                                1 - \sin(\pi (x_i -t)) & i \in [{{N+3}\over 2}, \cdots, N] \end{array}
                                               \right. ,
$$ 
with $x_{(N+1)/2} < 0 < x_{(N+3)/2}$. Let $v$ be the spectral approximation to Eq. (\ref{example1}) such that $v = (v_0, \cdots, v_N)^T$. Here $v_j$ denotes the grid function at $x = x_j$. Then we seek $v$ with the direct projection method such that 
\begin{eqnarray}
                    {{dv}\over {dt}} + {\bf D}\cdot v = \bf D \cdot \bf H,
\label{sp1}
\end{eqnarray}
where $d\over {dt}$ denotes the time integration operator applied to the discrete $v$. For the time integration, we use the 3rd order TVD Runge-Kutta scheme \cite{Shu}. The boundary condition is directly applied 
\begin{eqnarray}
              v_0 = -\sin(\pi(x_0 - t)),
\label{sp2}
\end{eqnarray}
after the final stage of the RK step at the given time $t$. 

Suppose that the numerical approximation of $v(x,t)$ is given in the form of $v(x,t) = w(x-t) + z(x)$ where $w(x-t)$ and $z(x)$ are the time-dependent and time-independent parts of $v(x,t)$ respectively. With the method of line, Eq. (\ref{sp1}) becomes
$$
            {{d w}\over {dt}} + {\bf D}\cdot (w + z) = \bf D \cdot \bf H, 
$$ 
and it is clear that 
$$
        {{d w}\over {dt}} + {\bf D}\cdot w = 0, \quad {\bf D} \cdot z = \bf D \cdot \bf H. 
$$
The boundary condition is given by 
$$
      v(-1,t) = u(-1,t) = -\sin(\pi(-1 - t)) + \int^{-1}_{-1} \delta(x) dx = -\sin(\pi(-1-t)),  
$$
that is, $w(-1,t) = -\sin(\pi(-1-t))$ and $z(-1) = 0$. Thus we know that $w(x,t)$ is the spectral collocation approximation of 
$ w_t + w_x = 0$ because its exact solution is given by $w(x,t) = -\sin(\pi (x-t))$. The time-independent solution $z(x)$ is then obtained from by solving ${\bf D} \cdot z = \bf D \cdot \bf H$. Since the boundary value of $z(x)$ is given as $z(-1) = 0$, $z(x)$ is given by 
$$
                    {\tilde {\bf D}} \cdot ({\tilde z} - {\tilde H}) = 0,
$$
where the superscript $\tilde{}$ denotes the inner matrix or vector without the first column and row. $\bf D$ is singular, but ${\tilde {\bf D}}$ is non-singular. Thus $z(x) = H(x)$ exactly. Here note that such exactness is obtained only at the collocation points. Due to the uniqueness of the solution of Eq. (\ref{sp1}), $v(x,t) = w(x,t) + z(x)$ indeed. The error function $E(x,t)$ is given by 
$E(x,t) = u(x,t) - v(x,t) = -\sin(\pi (x-t)) + H(x) - (w(x,t) + z(x)) = -\sin(\pi (x-t)) - w(x,t)$. Thus we know that the error is solely determined by the homogeneous solution $w(x,t)$ which is smooth function for $\forall t > 0$. Thus, the error decays spectrally. 

Since the initial condition $v(x,0)$ is simply given by $v(x,0) = -\sin(\pi(x-t))$, there exist the Gibbs oscillations due to the initial discrepancy of the initial condition. To see how the initial discrepancy changes with time, $t>0$, let $e(x,t) = v(x,t) - v^e(x,t)$ where $v^e(x,t)$ is the numerical solution with the {\it exact} initial condition. By definition, $e(x,0) \ne 0$. Since the boundary condition is the same for both $v$ and $v^e$, we know that $e(-1,t) = 0$ for $\forall t > 0$.  Since both $v$ and $v^e$ are obtained from the same equation, we obtain
$$
             {{d e}\over {dt}} + {\bf D}\cdot e = 0.
$$
Since $e(-1,t) = 0$ for $\forall t \ge 0$, $e(x,t) = 0$ for $t > 2$. Thus the initial Gibbs oscillations due to the discrepancy of the initial condition do not affect the solution behavior after a finite time $t$.

{\em Finite difference scheme:} For the comparison, we consider the Lax-Wendroff scheme to solve Eq. (\ref{example1}) with the direct projection method. Suppose that the finite difference solution is sought with the evenly spaced grids for a given $N$ such that 
$$
         x_j = x_0 + j h, \quad x_0 = -1, \quad h \equiv  \Delta x =  2/N. 
$$
Let $D^+$, $D^-$ and $D^0$ the forward, backward and central difference operators defined as \cite{Gustafsson}
$$
                  h D^+ v_j = v_{j+1} - v_j, \quad hD^-v_j = v_{j}-v_{j-1}, \quad 2hD^0 v_j = v_{j+1} - v_{j-1}.
$$
We also use the direct projection method for the approximation of the $\delta$-function as 
$$
            \delta_N(x) = {{d H_N(x)}\over dx},
$$
where $d\over {dx}$ is now the central difference operator $D^0$ with the Lax-Wendroff scheme such that 
\begin{eqnarray}
                 {{dH_N(x)}\over{dx}} \rightarrow D^0H_N(x) = {{H_N(x_{j+1})-H_N(x_{j-1})}\over{2h}},
\label{center}
\end{eqnarray}
where $H_N(x)$ is the grid function with the finite differencing grid of $H_0(x)$. This definition of $\delta_N(x)$ satisfies the definition of the $\delta$-function almost everywhere in the domain. For $u(x) = |sgn(x)|$, $\delta_N(x)$ yields for even $N$,
$$
      {d\over{dx}} \int^x_{-1} \delta(s)u(s) ds 
     \rightarrow ({\bf D}\cdot{\bf H}_N)_j u(x_j) = ({\bf D}\cdot{\bf H})_j = \left\{ 
                                                                                  \begin{array}{cc} {1\over{2h}}, & x_j = -\Delta x, \Delta x \\
                                                                                 0, & \mbox{otherwise} \end{array} \right. .
$$
Then the Lax-Wendroff scheme is given by 
$$
                  v^{n+1}_j = v^n_j + dt D^0 v^n_j + {{{dt}^2}\over{2}} D^+D^-v^n_j + D^0 H_N,
$$ 
where $dt$ is the time step and we choose $dt$ as small as the stability is guaranteed. 

To measure the errors we define the discrete $L_2$ and $L_\infty$ errors as 
$$
              L_2 ~Error = \sqrt{{1\over {N+1}}\sum_{i=0}^N (v(x_i,t) - u(x_i,t))^2},
$$
and 
$$  
            L_\infty ~Error = \max_{i\in [0,\cdots,N]}|v(x_i,t) - u(x_i,t)|.
$$

\begin{figure}[ht]
\begin{center}
\includegraphics[width=2.5in]{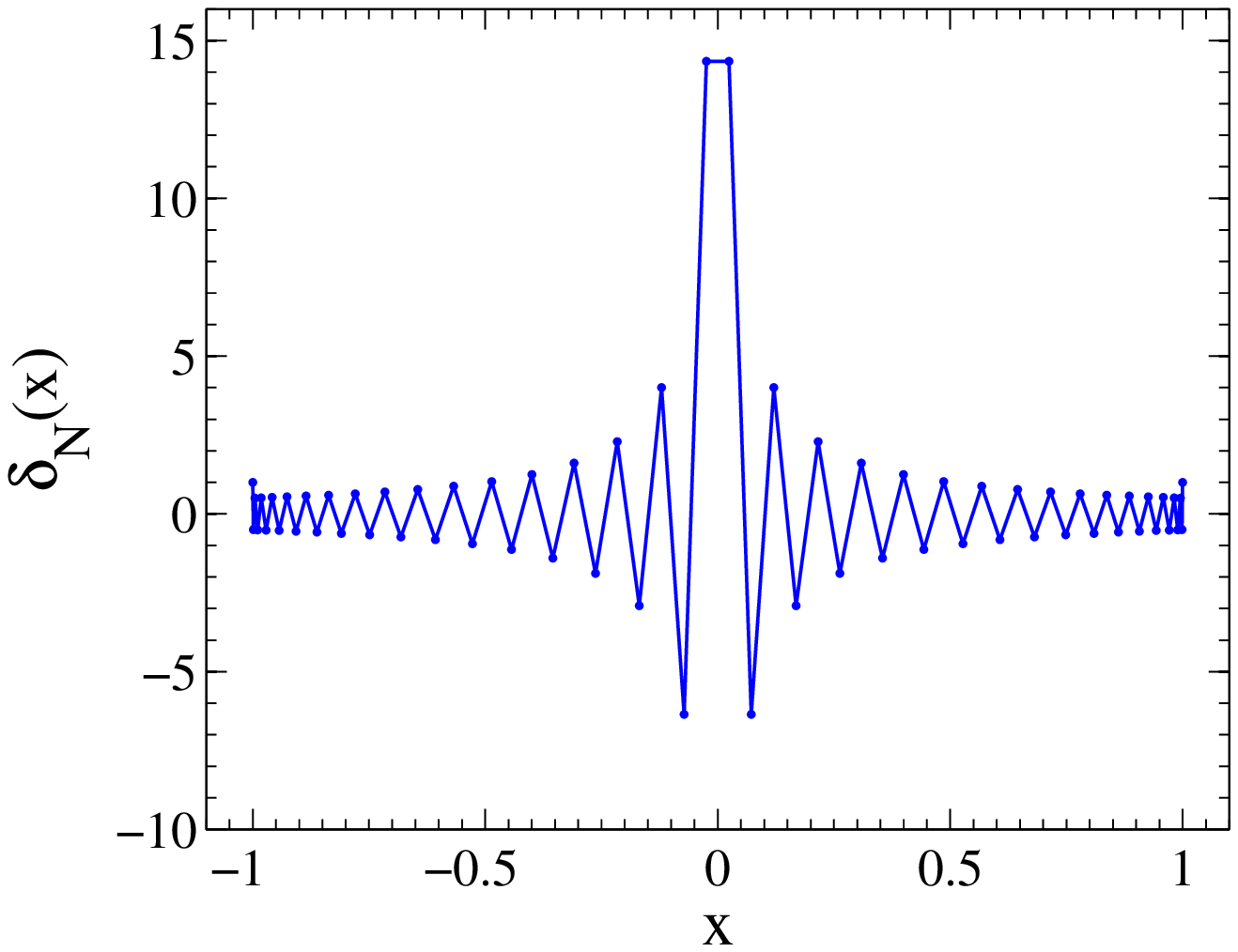}
\includegraphics[width=2.5in]{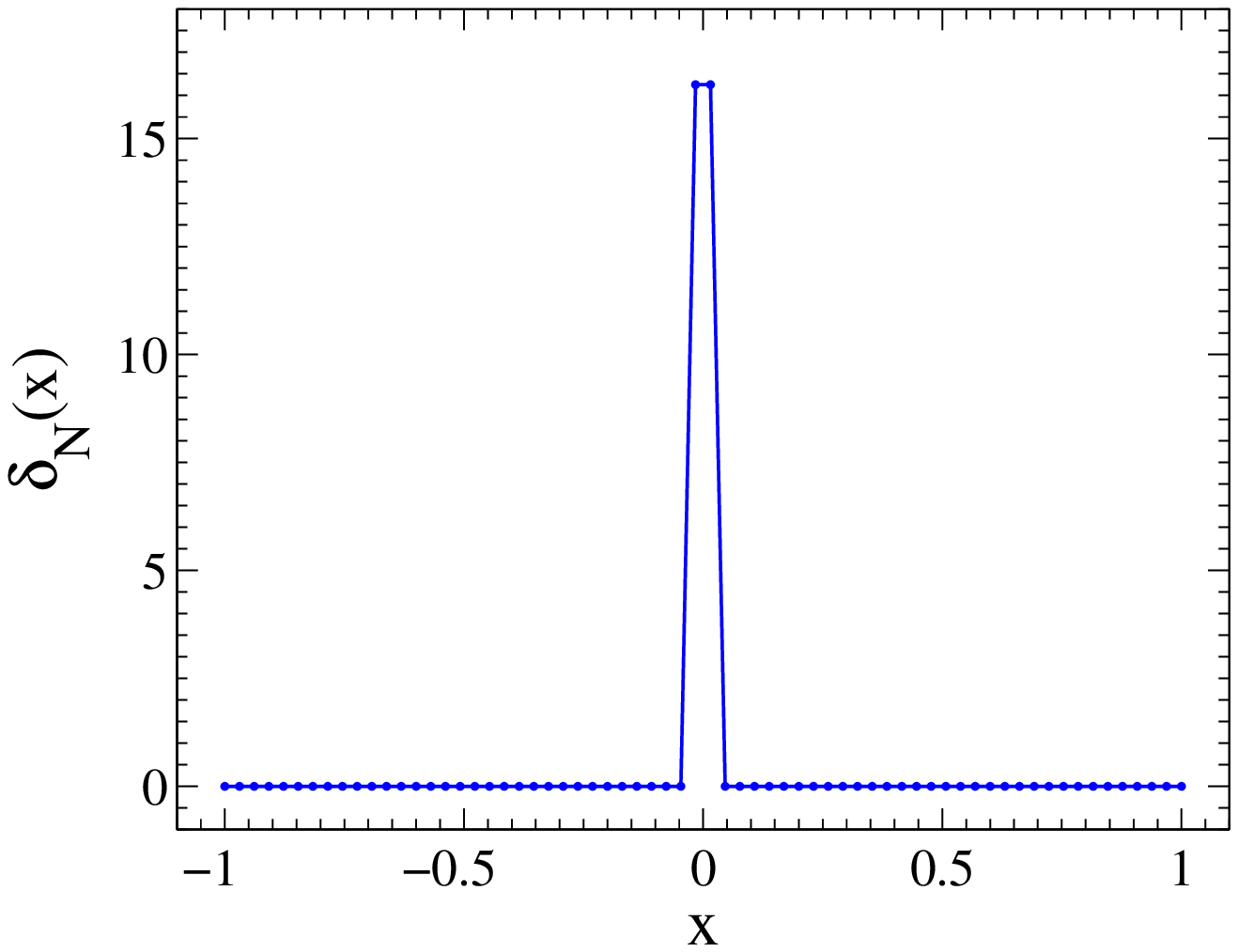}
\end{center}
\caption{{\it Left: The direct spectral projection approximation of $\delta(x)$ on the Chebyshev Gauss-Lobatto collocation points. Right: The direct central difference approximation of $\delta(x)$ on the evenly spaced grids with the central difference operator $D^0$ for $N = 66$.}} 
\label{figure1}
\end{figure}

Figure \ref{figure1} shows the approximations $\delta_N$ of $\delta(x)$ with the spectral method given in Eq. (\ref{SPdelta}) with $N = 65$ (left) and the centered difference scheme given in Eq. (\ref{center}) with $N = 66$ (right). As shown in the figures, the direct projection approximation of $\delta(x)$ with the Chebyshev method is highly oscillatory over the whole domain. Note again that $N$ is assumed to be odd to avoid any ambiguity at $x = c = 0$  for the spectral method. The approximation with the centered difference method, however, yields the approximation without any oscillations unlike the spectral approximation. The figure also shows that the $\delta$-function is highly localized around the singularity at $x = 0$ with the finite difference method. The oscillations shown in the spectral approximation is simply, the Gibbs phenomenon. It is also well known that the spectral solution is highly oscillatory if there is a local jump discontinuity such as the $\delta$-function. As mentioned earlier, in \cite{Jung} it has been shown that the direct projection method yields an accurate solution and spectral accuracy can be also achieved for certain types of PDEs although the approximation of the $\delta$-function obtained by this method is highly oscillatory.

\begin{figure}[ht]
\begin{center}
\includegraphics[width=2.7in]{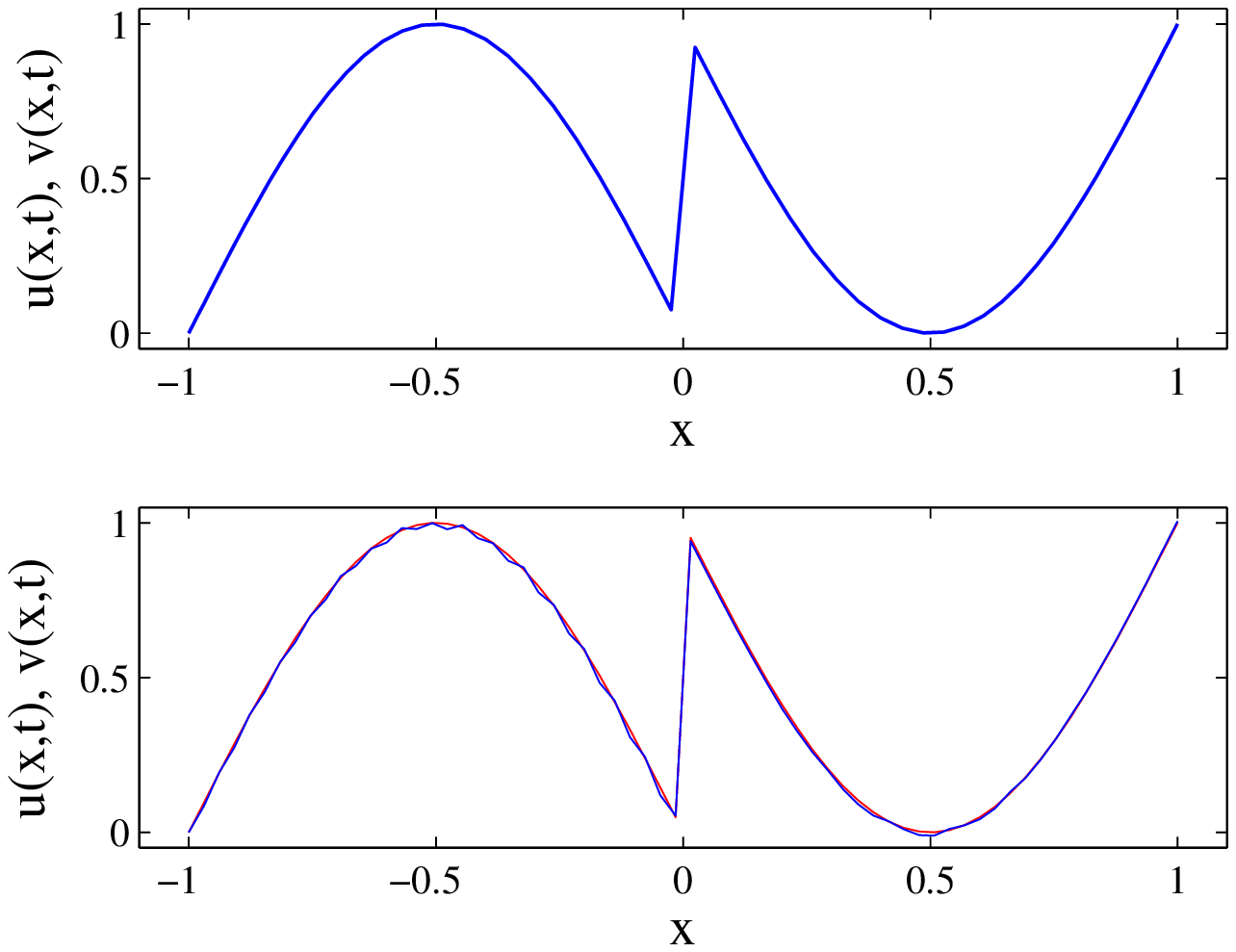}
\includegraphics[width=2.7in]{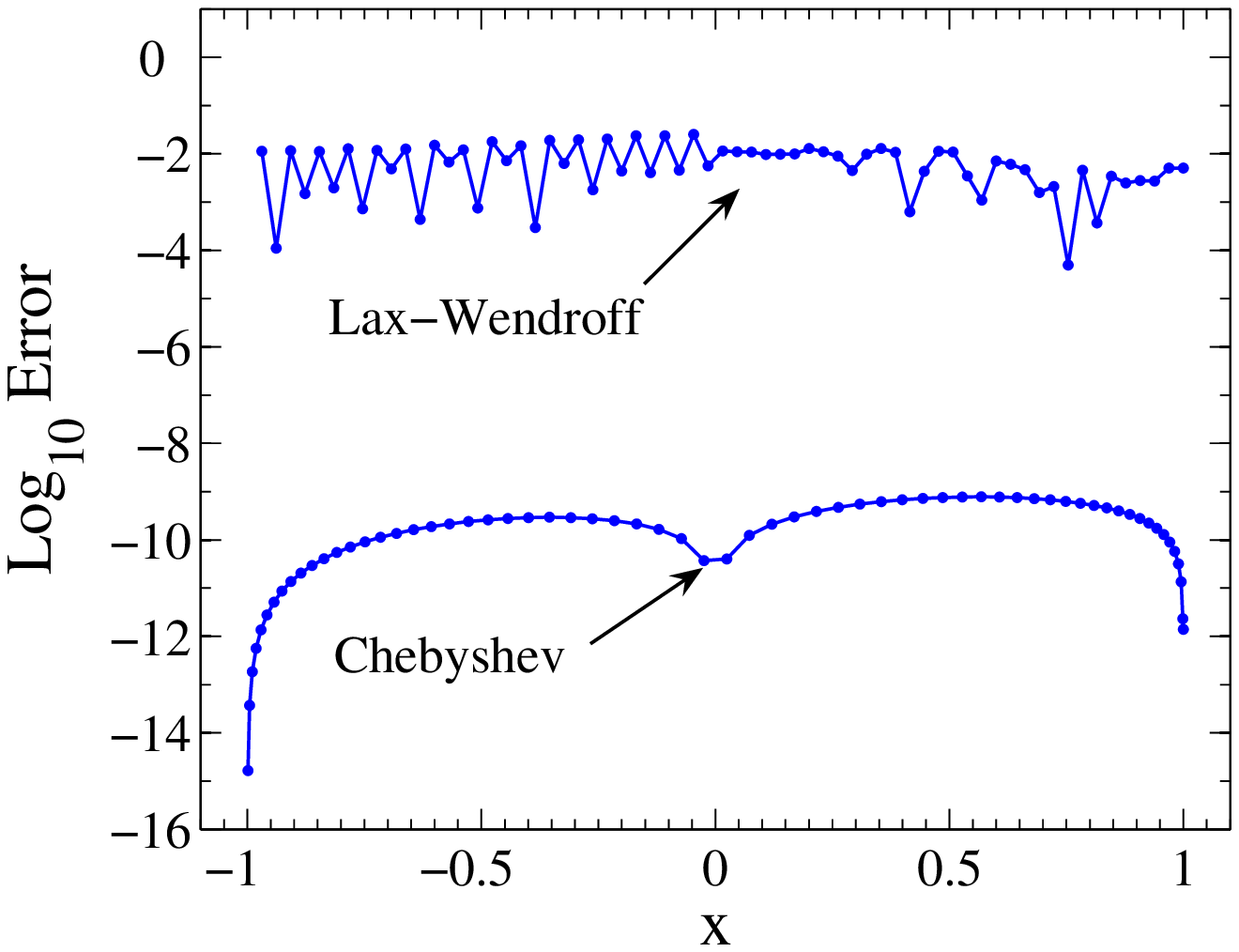}
\end{center}
\caption{{\it Left: Solutions at $t = 10$ with the Chebyshev spectral (top) and Lax-Wendroff (bottom) methods. The red and blue lines represent the exact and the numerical solution, respectively. Note that the exact and numerical solutions are not distinguishable for the Chebyshev spectral method. Right: Pointwise errors $Log_{10}|u(x,t) - v(x,t)|$ versus $x$ with the Lax-Wendroff method and Chebyshev spectral methods.} } 
\label{figure2}
\end{figure}

Figure \ref{figure2} shows the solutions and pointwise errors at $t = 10$. The left figure of Figure \ref{figure2} shows the solutions with the Chebyshev spectral (top) and Lax-Wendroff (bottom) methods. The red and blue lines represent the exact solution and the approximation, respectively. For the spectral approximation, the exact solution and the approximation are not distinguishable. There are no Gibbs oscillations observed in the spectral approximation. It seems that the direct spectral approximation of the $\delta$-function yields a  cancellation of these oscillations. In \cite{Jung}, it has been discussed that such cancellation is due to the consistent formulation to the given PDE. For the approximation with the Lax-Wendroff method, small oscillations are seen around the singularity at $x = 0$. In the right figure of Figure \ref{figure2}, the pointwise errors versus $x$ are given. The figure shows that the Chebyshev spectral method yields very accurate result over the whole domain without any indication of the degradation of convergence order. Indeed, with the Chebyshev spectral method, the spectral accuracy and uniform convergence are recovered even with the local jump discontinuity. Contrary to the spectral approximation, the approximation with the Lax-Wendroff method only shows the low order accuracy.   

\begin{figure}[ht]
\begin{center}
\includegraphics[width=2.5in]{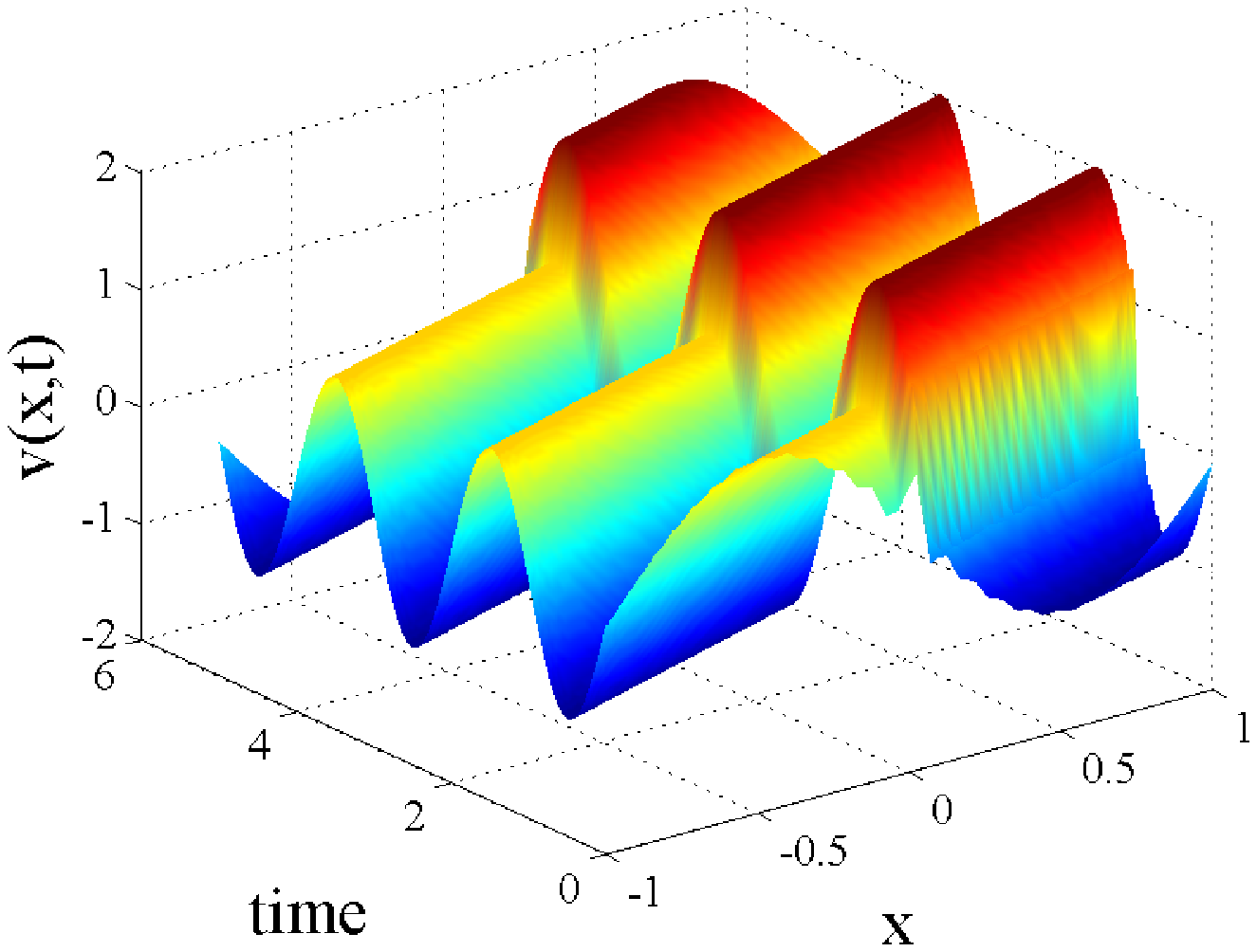}
\includegraphics[width=2.5in]{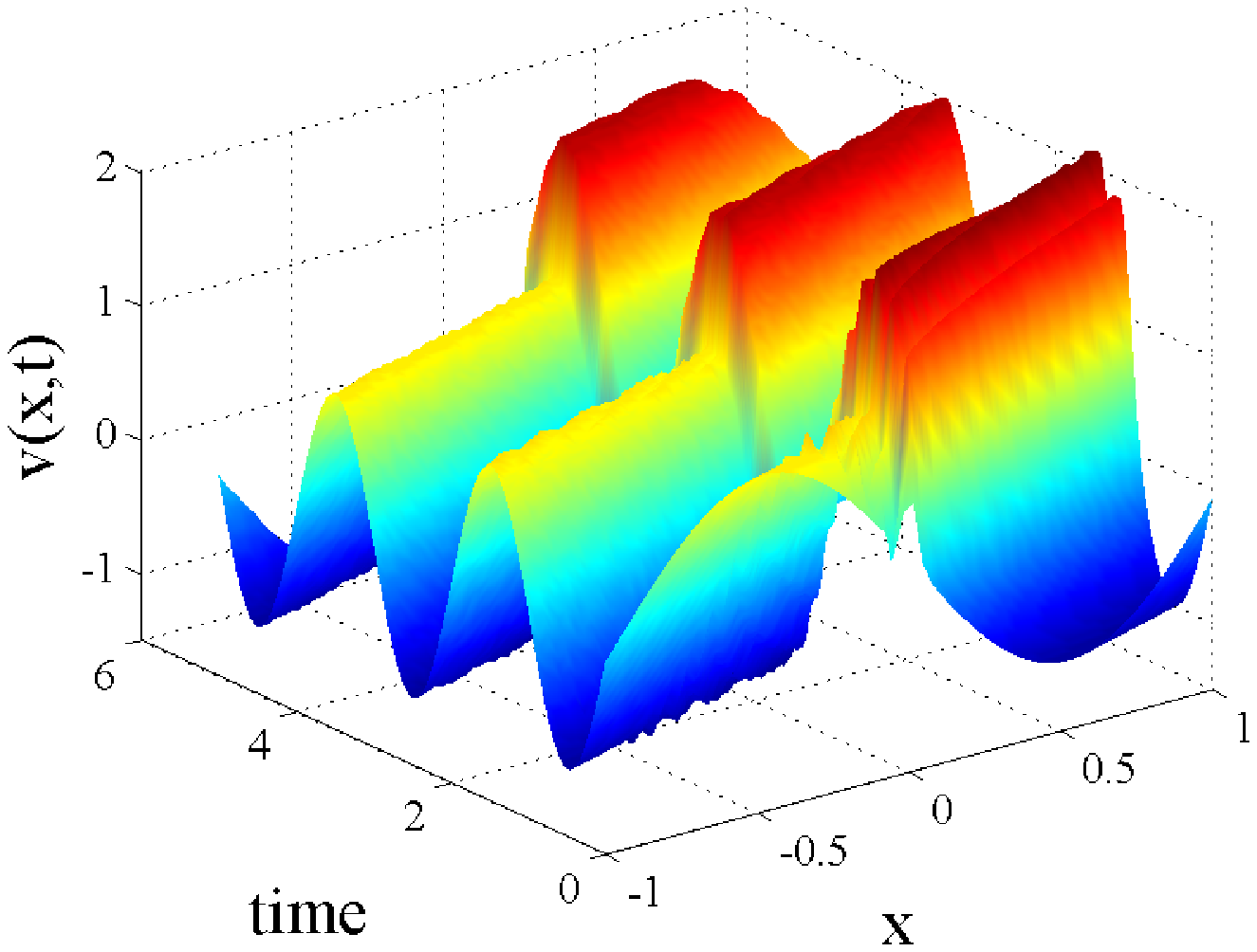}
\end{center}
\caption{{\it Trace of solutions with time with the Chebyshev spectral (left) and Lax-Wendroff methods (right).}} 
\label{figure3}
\end{figure}

Figure \ref{figure3} shows the trace of approximations with time with the Chebyshev spectral method (left) and the Lax-Wendroff method (right). There are small oscillations in the beginning time stages with both methods. These oscillations are due to the inconsistent initial condition. These oscillations with the Chebyshev spectral method, however, leave the domain quickly as time goes on while they remain with the Lax-Wendroff method. The right figure for the Lax-Wendroff method clearly shows that such oscillations exist with all time $t \in (0, 5]$ while no significant oscillations are seen in the left figure for the Chebyshev spectral method even after $t>T$. Here $T$ is the time for the initial Gibbs oscillations to leave the domain and it is about $T \sim 1$ as the wave speed is $1$.

Table 1 shows the $L_2$ and $L_\infty$ errors of the solution for various approximations of the $\delta$-function. In the table, LW-Direct, LW-Gaussian, SP-Direct and SP-Gaussian represent the Lax-Wendroff method with the direct and Gaussian approximations of the $\delta$-function, and the Chebyshev spectral method with the direct and Gaussian approximations of the $\delta$-function, respectively. As shown in the table, the best result is obtained with the Chebyshev spectral method with the direct approximation of the $\delta$-function. The table also shows that the direct projection method yields a slight better result for the Lax-Wendroff method. 
\begin{table}[ht]
\begin{center}
\caption{\it{$L_2$ and $L_\infty$ errors with various approximations of $\delta(x)$ at $t = 10$ for Eq. (\ref{example1}). }}
\end{center}
\begin{center}
\begin{tabular}{lrr}
\hline
   Method   & $L_2$ error & $L_\infty$ error  \\ \hline
\mbox{LW-Direct} & 1.02(-2) & 2.52(-2)  \\
\mbox{LW-Gaussian} &4.52(-2)  &  2.58(-1)  \\
\mbox{SP-Direct} & 3.60(-10) & 7.77(-10)  \\
\mbox{SP-Gaussian} &4.04(-2)  &  2.33(-1)   \\
\hline
$(n) = 10^{n}$
\end{tabular}  \label{table1}
\end{center}
\end{table}

\section{Numerical approximation of Zerilli equation}

Now we consider numerical approximations of the inhomogeneous Zerilli equation with the spectral and finite difference methods. For both the spectral and finite difference methods, we use the direct projection method of the $\delta$-function. Here we note that the finite difference method used in this work is the explicit second-order centered finite difference method. High order finite difference methods for the Zerilli equations such as a  fourth-order convergent finite difference method, have been developed \cite{Lousto2}. The comparisons between the spectral and high order finite difference approximations for the Zerilli equations will be left for our future work. 

The governing equation is defined on the simple $1+1$ spacetime with time and space denoted by $t$ and $r^*$ respectively. 
The range of $r^*$, the tortoise coordinate is given by 
$$
                                         R^*_e \le r^* \le R^*_\infty,
$$
where $R^*_e$ is obtained close to the radius of the event horizon and $R^*_\infty \rightarrow \infty$. In practice, $R^*_e$ and $R^*_\infty$ can not be $-\infty$ and $\infty$ respectively. Let $R^*_e$ is taken at $r \sim 2M$ and $R^*_\infty$ at $r \gg 2M$. By definition, $R^*_e <0$ and $R^*_\infty >0$. The tortoise coordinate $r^*$ is linearly mapped into the Chebyshev Gauss-Lobatto collocation points $\{\xi_j\}_{j=0}^N$ given as 
\begin{eqnarray}
                        r^* = {{R^*_\infty - R^*_e}\over 2}\xi + {{R^*_\infty + R^*_e}\over 2},
\label{linmap}
\end{eqnarray}
where 
$$
                         \xi_j = -\cos(\pi j/N), \quad j = 0, \cdots, N. 
$$

\subsection{Derivative operator $D$}
Using the linear mapping of Eq. (\ref{linmap}), the derivative operator with respect to $r^*$, $D$ is given by 
$$
                       D:= {{\partial}\over{\partial r^* }} = {2\over {R^*_\infty - R^*_e}} {{\partial}\over{\partial \xi }} = 
                                                              {2\over {R^*_\infty - R^*_e}}D(\xi),
$$
where $D(\xi)$ is the differential operator defined on the Gauss-Lobatto collocation points $\{\xi_j\}_{j=0}^N$ given in Eq. (\ref{equation19a}).
The second derivative operator is given by
$$
            D^2 = D\cdot D =  \left({2\over {R^*_\infty - R^*_e}}\right)^2D^2(\xi).
$$
Then the Chebyshev spectral method for the black hole system Eq. (\ref{zerilli}) seeks an approximation for 
\begin{eqnarray}
                          {{d^2\psi}\over{dt^2}} = D^2 \psi - V_l \psi - S,
\label{system}
\end{eqnarray}
where $\psi$, $V_l$ and $S$ are the wave function, the potential of $l$ mode and the source term defined on $\{\xi_j\}_{j=0}^N$, respectively. Singular source terms $\delta(r,t)$ and $\delta'(r,t)$ are approximated with the direct projection method as 
$$
           \delta(r-R) = D \cdot H_{R}, \quad \delta'(r-R) = D\cdot \delta(r-R) = D^2\cdot H_{R},
$$
where $H$ is the Heaviside function defined on $\{\xi_j\}_{j=0}^N$. Here we note that $R$ is a function function of time and the Heaviside function $H_R$ is being shifted as time goes on. 

In our computation, the second derivative of $\psi$ is with respect to $r^*$ not $r$. Thus at each collocation point, $r^*$, the corresponding proper distance should be calculated to obtain the potential and singular source terms accordingly, for which the bi-section method is used. 

\subsection{Time stepping}
For the time-stepping, we adopt the second-order time stepping method 
$$
     \psi_{tt} = {{\psi(t+ \Delta t) - 2\psi(t) + \psi(t-\Delta t) }\over{\Delta t^2}}.
$$
This method is two-step method. Thus we need two initial conditions for the time marching. We can simply use 
$$
            \psi(r,0) = \psi(r,\Delta t), 
$$
or we can evolve Eq. (1) using the first order approximation for $\psi(r,\Delta t)$ \cite{Gustafsson}. 

\subsection{Initial condition}
The initial location of the point particle is $r = r_0<\infty$. We have to find the exact initial condition when the particle is located at finite $r = r_0$ initially. In this paper, we assume that 
$$
              \psi(r, t = 0) = 0. 
$$
This vanishing initial condition is not physical and inconsistent to Eq. (\ref{zerilli}) as it does not satisfy the governing equation. Due to the inconsistent treatment of the initial condition, the numerical approximations of $\psi$ in the beginning stages are highly oscillatory and it takes a certain time for these initial oscillations to leave the computational domain as we will see in the next sections.  

\subsection{Boundary conditions}
We impose the absorbing boundary condition at the boundaries $r^* = R^*_e$ and $r^* = R^*_\infty$. Since this is the 1D problem, the following boundary condition should be a perfectly absorbing boundary condition:
$$
                  \psi_t - \psi_{r^*} = 0, \quad\quad at \quad r^* = R^*_e,
$$ 
and 
$$
                  \psi_t + \psi_{r^*} = 0, \quad\quad at \quad r^* = R^*_\infty.
$$
Here we note that the derivative with the spectral method at the end points should not be obtained using the global derivative operator $D$ because there can be a discontinuity inside the domain. That is, the first order derivative term is not given as $D\cdot \psi|_{r= R_e}$. The no-flux boundary conditions should be only {\it local}. We use the upwind method at both ends such as  
$$
             {{\psi(r^*_0,t+\Delta t)-\psi(r^*_0,t)}\over{\Delta t}} = {{\psi(r^*_1,t)-\psi(r^*_0,t) }\over{r^*_1 - r^*_0}}, \quad\quad at 
                     \quad r^* = R^*_e,
$$ 
where $r^*_0 = R^*_e$ and 
$$
             {{\psi(r^*_0,t+\Delta t)-\psi(r^*_0,t)}\over{\Delta t}} = -{{\psi(r^*_N,t)-\psi(r^*_{N-1},t) }\over{r^*_N - r^*_{N-1}}}, \quad\quad at 
                     \quad r^* = R^*_\infty,
$$ 
where $r^*_0 = R^*_\infty$. 

\subsection{Spectral filtering methods}
We also adopt the spectral filtering method to minimize the possible non-physical high-frequency modes. The oscillations with the spectral method possibly found near the local jump discontinuity and also generated due to the inconsistent initial conditions, propagate through the whole domain. To reduce these non-physical modes, a filter function is applied to the solution. To apply the filter function, we transform the approximation found in the physical domain to the coefficient domain using Eq. (\ref{coefficient}). Then we obtain the filtered approximation as 
$$
                  u^\sigma_N(x_l,t) = \sum_{j=0}^N \sigma(j/N) {\hat u}_j T_j(x_l), 
$$
where $\sigma(l/N)$ is the filter function according to \cite{Vandervan}. The filtering method is equivalent to the spectral viscosity method \cite{Hesthaven}. For the current work, we use the exponential filter defined as
$$
                 \sigma(j/N) = \exp(\ln \epsilon (j/N)^p), \quad \forall j = 0, \cdots, N,
$$
where $\epsilon$ is a strictly positive real constant and $p$ is known as the filtering order. The filter function is constructed to satisfy the properties, $\sigma(0) = 1$, $\sigma^{(q)}(0) = 0$ for $q = 1, \cdots, p-1$, $\sigma(\pm 1) = 0$, and $\sigma(-\eta) = \sigma(\eta)$ \cite{Vandervan}. 
In principle, $u^\sigma_N(t) \rightarrow u_N(t)$ if $p \rightarrow \infty$. If $p \rightarrow 0^+$, $u^\sigma_N(t)$ becomes only 0{\it th} order approximation of $u_N(t)$. By the definition of the filter function $\sigma(j/N)$, $\sigma(1) \rightarrow 0$, for which the positive constant $\epsilon$ is usually chosen so that $\exp(\ln \epsilon)$ becomes about machine accuracy, $\epsilon_M$. For our work, we choose $\epsilon_M \sim 10^{-16}$. Using Eq. (\ref{coefficient}), we have 
$$
              u^\sigma_N(x_i) = \sum_{j=0}^N S_{ij} u_j, \quad \forall i = 0, \cdots, N,
$$ 
where $S_{ij}$ are 
$$
                         S_{ij} = {1\over {c_j}}\sum^N_{n = 0}{2\over{c_n N}}T_n(x_i)T_n(x_j)\exp(\ln \epsilon (n/N)^p).
$$

\subsection{Filtering for the finite difference method}
For the finite difference method, we also experimented with a filtering method. We simply convolved a Gaussian distribution with the source-term itself, yielding a smoother profile for it. This Gaussian filter approach has been used elsewhere \cite{pranesh2} in literature is a natural one to attempt in our context. Through experimentation, we realized that choosing the width of the Gaussian to be of the same scale as the spacing between grid points yields the best results. Figure \ref{figure14} depict sample results from performing such Gaussian filtering in the finite difference evolutions. These results show that the Gaussian filtering yield better results than those without the filtering. With the Gaussian convolution the filtered function $f^{filtered}_N$ for $f_N$ is given by  
$$
      f^{filtered}_N(x,t) = \int^\infty_{-\infty} f_N(y,t)\exp(-(x-y)^2/\sigma)dy,
$$
The Gaussian filter is applied for the source term. Since the source is only localized, the integration is done only over a certain number of grids, that is, for example at $i$th grid point, $x = x_i$, the integration on the grids is done over $[x_{i-n_p}, x_{i+n_p}]$. Thus totally $2n_p +1$ grid points are used for the integration. According to our numerical experiments, the best parameters for the Gaussian kernel for various $N$ are 
$(N,n_p,\sigma) = (32, 2, 8), (64, 6, 8), (128, 32, 16)$, and $(1024, 32, 32)$. These values are used for the filtering for the finite difference method in the following sections. 

\subsection{Numerical results}

For the numerical experiments, we use $M = 1$, $m_0 = 10^{-10}$, $R^*_e = -30$, $R^*_\infty = 30, 130$, and $l = 2$. For the time integration we use $dt = 5\times 10^{-5}$ with the final time $T_{final} = 300, 600,$ and $2400$.  For the spectral case, filtering orders in the range of $p = [32, 64]$ were used. For the finite difference method, the filtering parameters given in Section 4.6 are used. For the comparison of the spectral method with the finite difference method, we focused on accuracy and convergence of the ringdown profile with time and $N$. Since the current work is based on the spectral method with the single domain, the finite difference method has faster computing time than the spectral method. The multi-domain spectral method will be considered in our future work to increase the computing speed. 

\begin{figure}[ht]
\begin{center}
\includegraphics[width=2.7in]{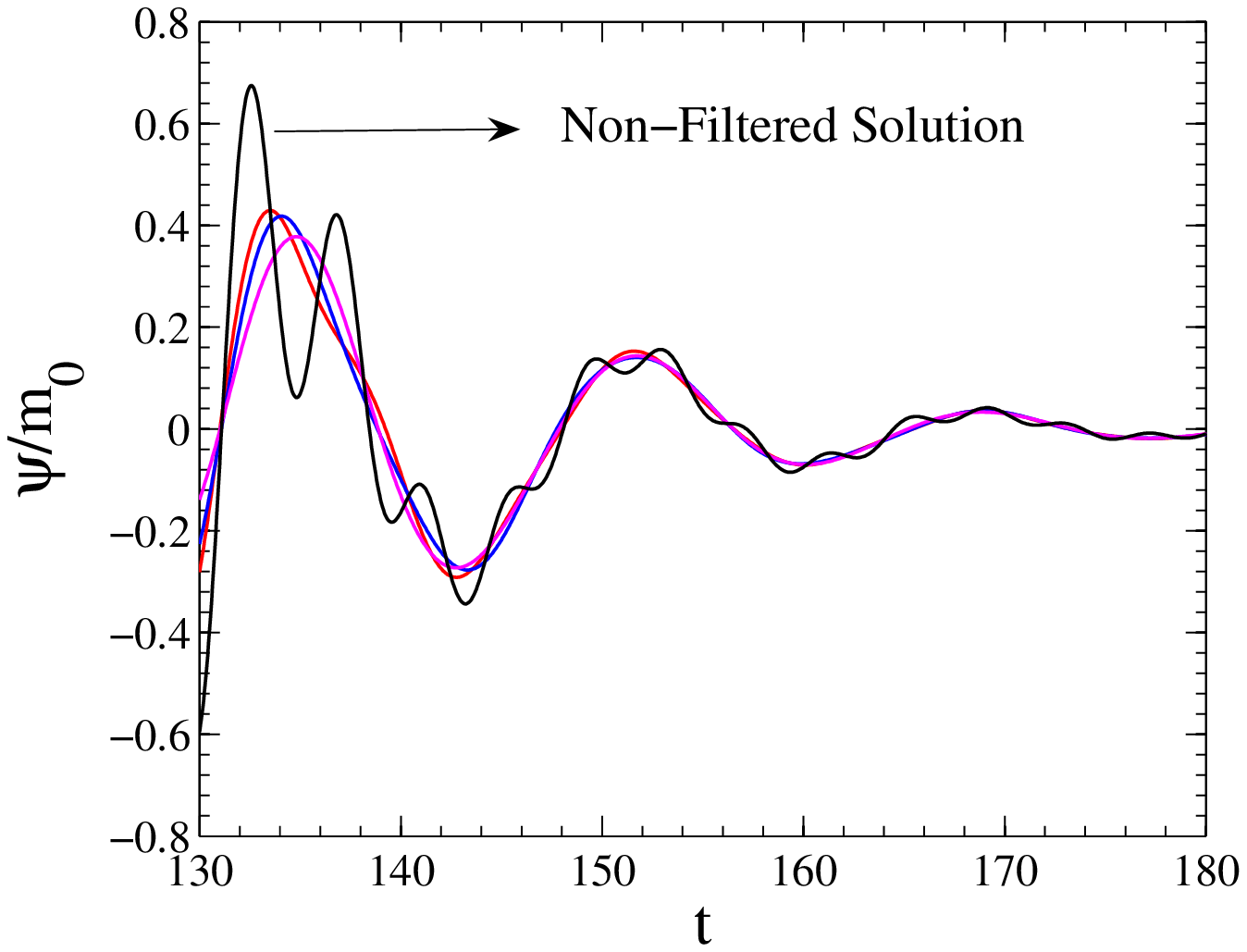}
\includegraphics[width=2.7in]{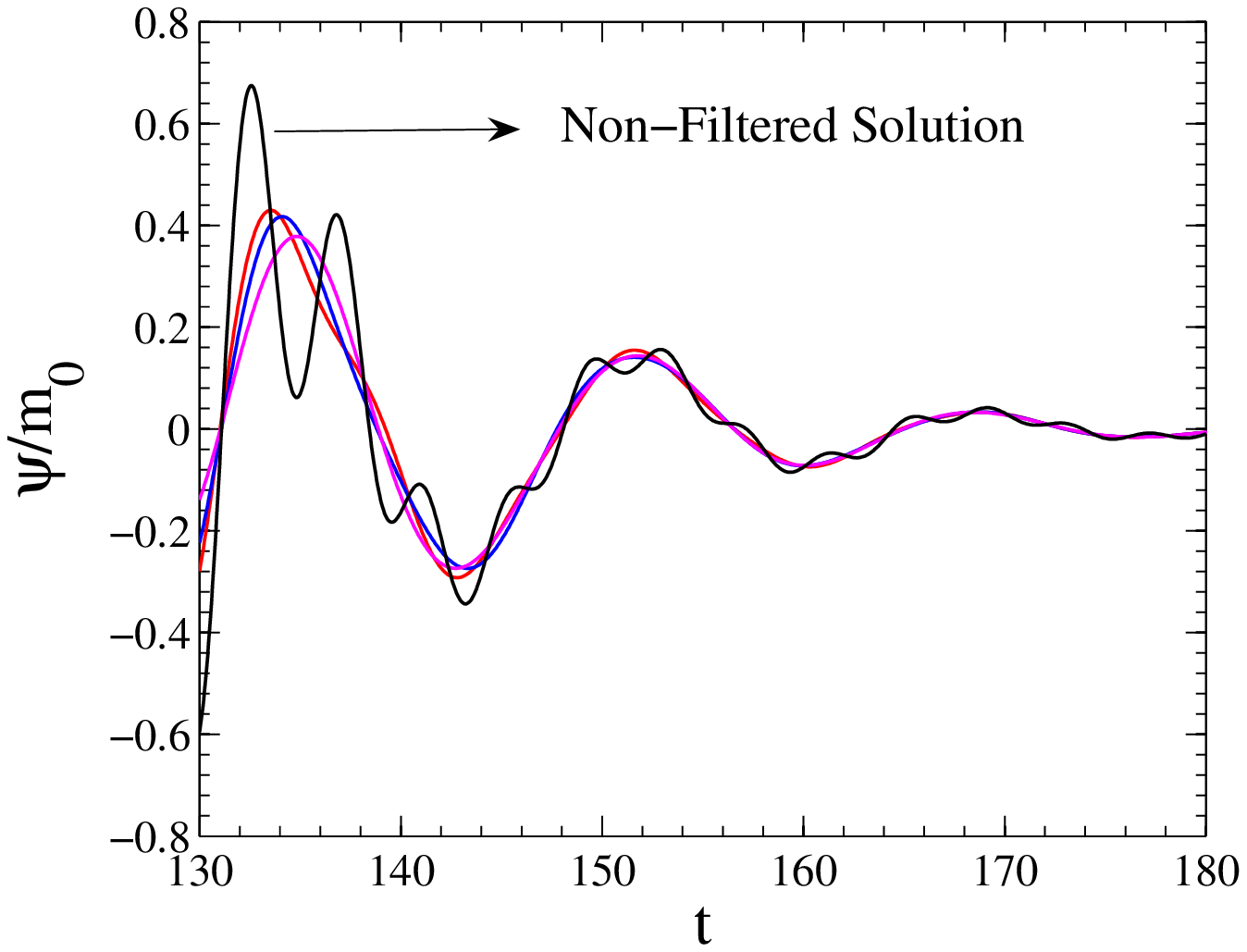}
\end{center}
\caption{\small{\it{$\psi/m_0$ versus $t$ with the spectral method for $p = 64$ (red), $p = 48$ (blue), $p = 32$ (magenta) and unfiltered method (black) for $t \in [130, 180]$. Left: $t_{filter} = 0$. Right: $t_{filter} = 10$.  
}}} 
\label{figure6}
\end{figure}

Figure \ref{figure6} shows the waveform $\psi/m_0$ of unfiltered and filtered spectral solutions for $130\le t \le 180$ with $N = 128$. For filtered solutions, $p = 64, 48,$ and $32$ are used. The left figure shows the case of filter on-set time, $t_{filter} = 0$ and the right of $t_{filter} = 10$. The oscillations shown in the unfiltered solution are considerably reduced in the filtered solutions, and the filtered solutions with different $p$ agree well. It is also observed that the filter on-set time does not affect the solution in the ringdown phase if $t_{filter} \le t \sim 130$. 

\begin{figure}[ht]
\begin{center}
\includegraphics[width=2.7in]{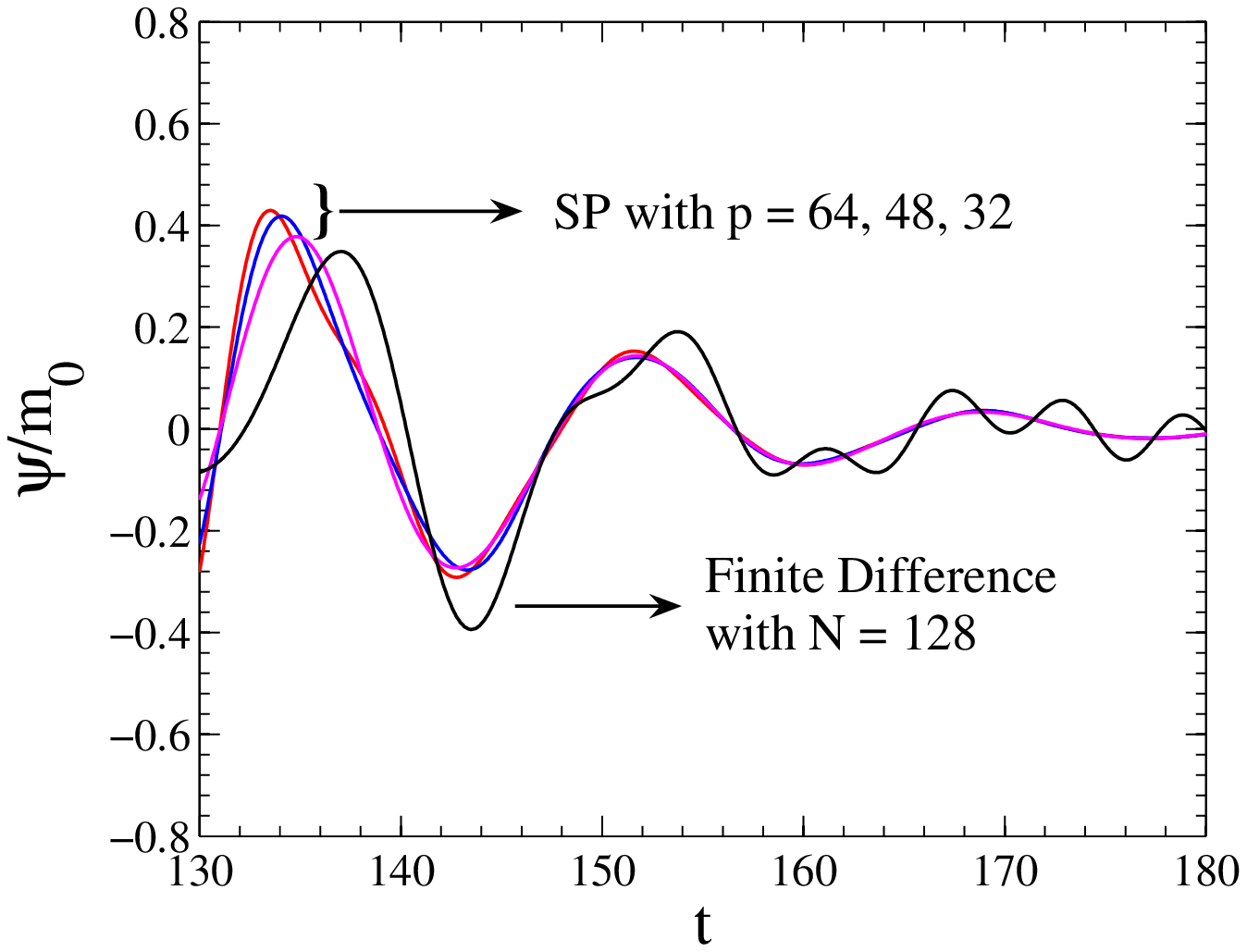}
\includegraphics[width=2.7in]{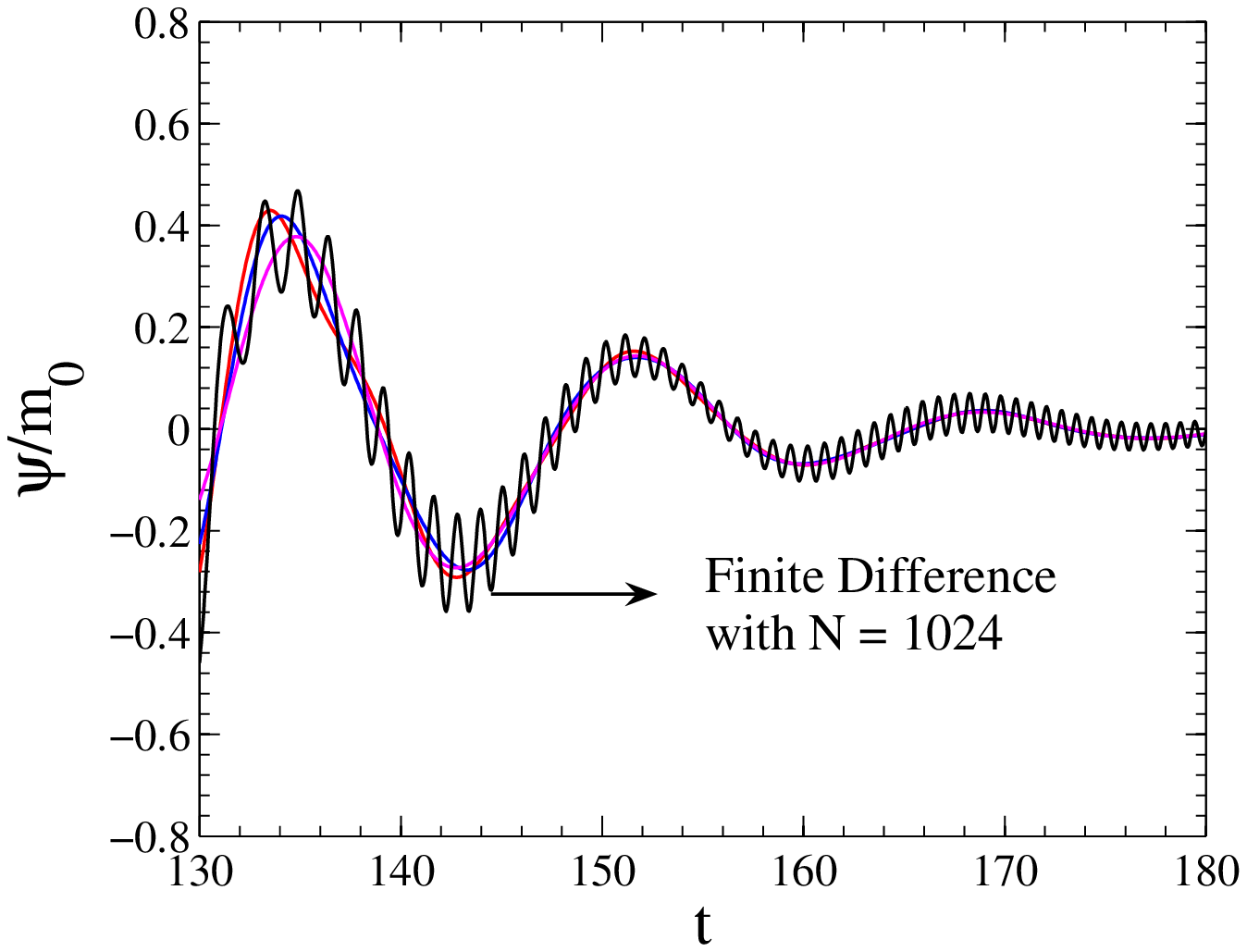}
\end{center}
\caption{\small{\it{$\psi/m_0$ versus $t$ with the spectral and finite difference methods. For the spectral method, $p = 64$ (red), $p = 48$ (blue), and $p = 32$ (magenta) are used with $N = 128$ and $t_{filter} = 0$ for both plots. For the finite difference method, $N = 128$ (left) and $N = 1024$ (right) are used.  
}}} 
\label{figure7}
\end{figure}

Figure \ref{figure7} shows the waveforms $\psi/m_0$ of filtered spectral solutions and those of the finite difference method. $N = 128$ is used for the spectral method and $N = 128$ (left) and $N = 1024$ (right) are used for the finite difference method. No filtering is used for the finite difference method. The figures show that the waveforms are highly oscillatory with the finite difference method and also show that more oscillatory waveforms are found with larger $N$. Furthermore the figures show that the global behavior of the finite difference solution with large $N = 1024$ agree with that of the filtered spectral solutions better than with $N = 128$. This implies that the spectral method can yield more accurate solution with smaller $N$ than the finite difference method with the proper choice of $p$.   

Based on the solutions obtained, we conducted a Fourier analysis to find the period $T$ and the frequency $\omega$ of 
the ringdown waveforms. Table \ref{table2} shows the period and frequency obtained with the Fourier analysis. In the 
table $CD$ denotes the finite difference method with the central differencing, $SP$ the spectral method and $p$ the 
spectral filtering orders. The period computed based on the frequency domain approach is well known as $T \sim 16.815$. 
The table clearly shows that the periods of the ringdown waveforms with the spectral method for $N = 32$ to $N = 128$ 
and with the finite difference method for $N = 128$ and $N = 1024$ are about $T \sim 16.8$. The finite difference method 
with smaller $N$ such as $N = 32, 48,$ and $64$, however, yields inaccurate results of $T \sim 18.7$ and $T \sim 15.3$. 
The table also shows that the spectral method yields an accurate period even with smaller $N$ such as $N = 32$. 

\begin{table}[ht]
\caption{\it{The period $T$ and frequency $\omega$ of the ringdown of the waveform $\psi/m_0$ for $l = 2$.}}
\begin{small}
\begin{tabular}{l|cc|cc}
\hline
Method  & $N$         & $p$            & $\omega$              & $T$   \\ \hline
CD      & $32$        &                & $0.33649906435356$    & $18.672$      \\
CD      & $48$        &                & $0.33649906435356$    & $18.672$      \\
CD      & $64$        &                & $0.41127663420991$    & $15.277$      \\
CD      & $128$       &                & $0.37388784928174$    & $16.805$      \\
CD      & $1024$      &                & $0.37388784928174$    & $16.805$      \\ \hline
SP      & $32$        &                & $0.37388784928174$    & $16.805$      \\
SP      & $48$        &                & $0.37388784928174$    & $16.805$      \\
SP      & $64$        &                & $0.37388784928174$    & $16.805$      \\
SP      & $128$       &                & $0.37717970828492$    & $16.658$      \\
SP      & $128$       & $64$           & $0.37388784928174$    & $16.805$      \\
SP      & $128$       & $48$           & $0.37388784928174$    & $16.805$      \\
SP      & $128$       & $32$           & $0.37388784928174$    & $16.805$      \\
\hline
\end{tabular}  \label{table2}
\end{small}
\end{table}

\begin{figure}[ht]
\begin{center}
\includegraphics[width=2.7in]{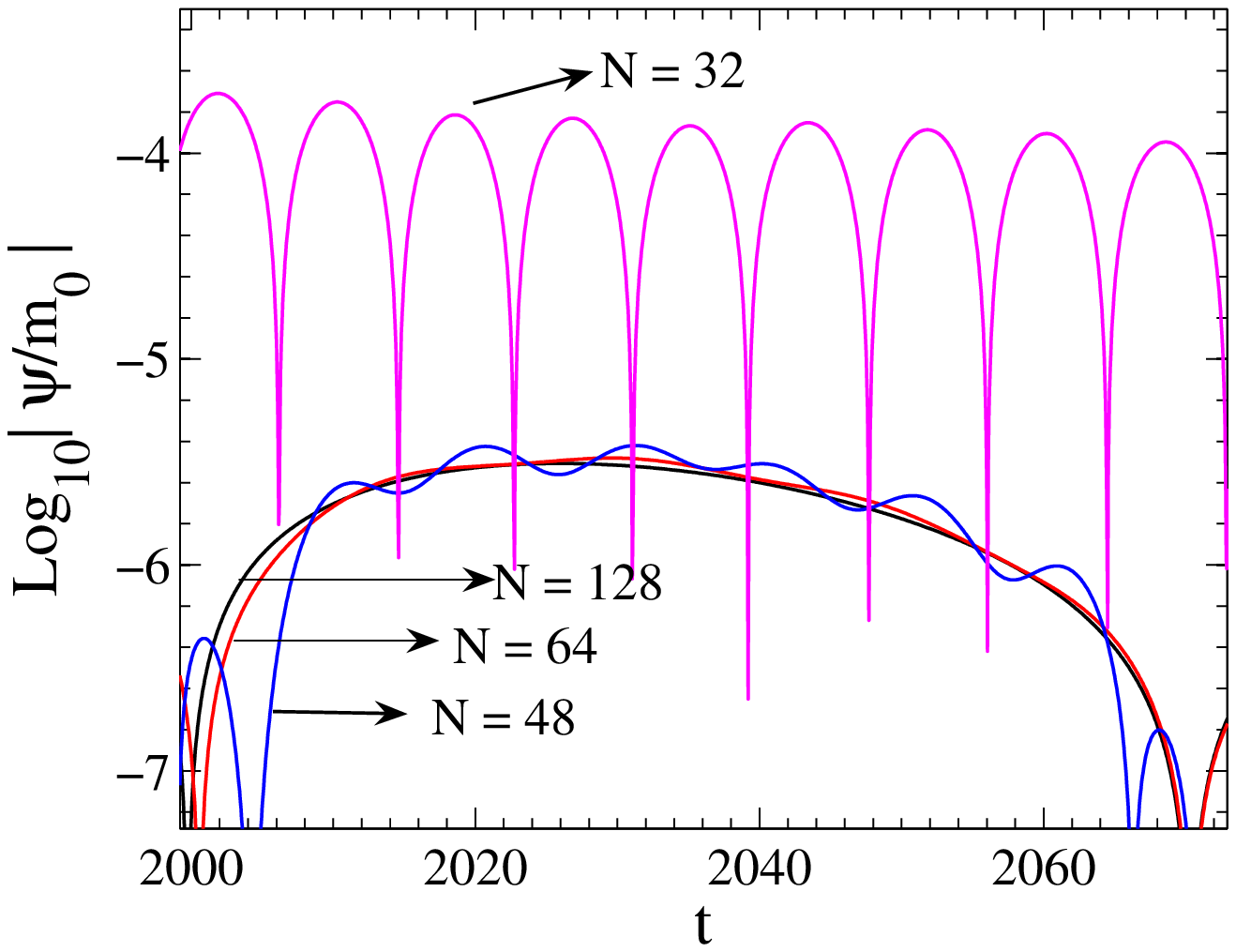}
\includegraphics[width=2.7in]{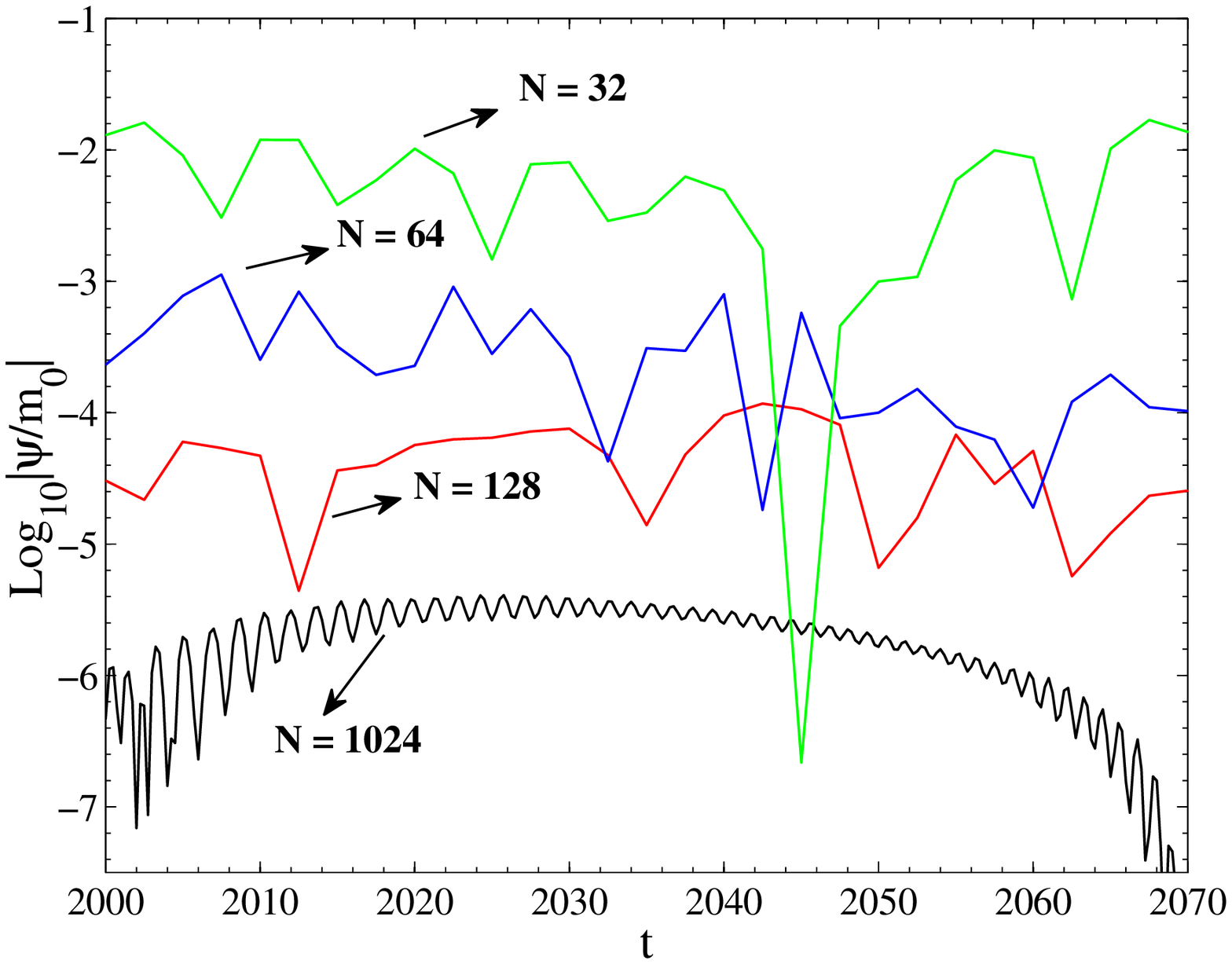}
\end{center}
\caption{\small{\it{$Log_{10}|\psi/m_0|$ versus $t$. $t \in [2000, 2070]$.  
Left: Unfiltered spectral approximations with $N = 128$(black), $64$(red), $48$(blue), and $32$(magenta). 
Right: Filtered finite difference approximations with $N = 1024$ (black), $128$ (red), $64$ (blue), and $32$ (green). 
}}} 
\label{figure14}
\end{figure}

Figure \ref{figure14} shows the waveform in logarithmic scale for $t \in [2000,2070]$. The left figure shows the unfiltered spectral approximations with $N = 128$  (black), $N = 64$ (red), $N = 48$ (blue), and $N = 32$ (magenta). The right figure shows the filtered finite difference approximations with $N = 1024$ (black), $N = 128$ (red),$N = 64$ (blue) and $N = 32$ (green). The figures clearly show that the spectral approximation yields a fast convergent behavior as $N$ is increased, without any filtering method applied. The finite difference approximations, on the other hand, yield only a slow convergence as the right figure indicates. Moreover, large $N$ such as $N = 1024$ is needed for the finite difference method to achieve comparable accuracy as the one obtained by the unfiltered spectral solution with $N = 64$. It is worth noting that the convergence with the finite difference method is possible only if the Gaussian filtering is used.

\section{Summary \& Outlook}
In this work, the inhomogeneous Zerilli equation has been solved numerically using both the Chebyshev spectral collocation method and the 2nd order explicit finite difference method to investigate the radial-infall of a point particle into a Schwarzschild black hole. The infalling object is considered as a point source and consequently singular source terms appear in the governing equations. For the approximation of the singular source terms, we use the direct derivative projection of the Heaviside function on the collocation points. Such an approach yields very oscillatory approximations for the $\delta$-function and its derivative. Despite these oscillations, the spectral solution is obtained in an accurate way as the direct derivative projection is consistently formulated with the given PDEs. A simple wave equation in $1+1$D spacetime shows that the spectral method recovers the exponential convergence even with the local jump discontinuities. The same technique is applied to solve the Zerilli equation and various numerical results have been presented. These results are also compared with those from the second-order explicit finite difference method. The spectral approximation with large $N$ shows small oscillatory behavior after the effect of the initial oscillations is minimal. To reduce these unphysical oscillations, the spectral filtering method is applied. A proper order $p$ of filtering, yields a smooth waveform with these oscillations considerably reduced. The finite difference method needs large $N$ to obtain the approximation with the same accuracy of the spectral method. Therefore, the spectral method yields accurate results with much smaller $N$ and can be efficiently used for the approximation of these types of problems. 

Our future work will center around the development of more efficient spectral methods such as those with the domain decomposition techniques and also adaptive filtering. These will further improve the accuracy and the computational efficiency of the solution. For the multi-domain spectral method or the discontinuous Galerkin method, it is possible to place the singular source term at the domain or element interface so that the singular source term plays a role only as a boundary or interface condition. In this approach, the singular source term is involved in the given PDEs only in a weak sense. For the spectral multi-domain penalty method, the penalty parameter can be also exploited to enhance both stability and accuracy associated with the singular source term. In general, however, the location of the singular source term varies with time and it would be more convenient to allow the singular source term to exist inside the sub-domain or element, in particular when the range of the possible locational variation is large. 

Finally, we will also investigate more general collisions of black holes (for example, in 2 or 3 dimensions, with rotation, etc.) and compare with the results from the frequency domain method. 

\section{Acknowledgements}
JHJ was supported by the NSF under Grant No. DMS-0608844. GK and IN acknowledge research support from the UMD College of Engineering RSIF and the Physics department.

\end{document}